\documentclass[preprint,superscriptaddress,nofootinbib,amsmath,amssymb,prd]{revtex4-2}
\usepackage{graphicx}
\usepackage{hyperref}
\usepackage{physics}
\usepackage{xcolor}
\DeclareUnicodeCharacter{2212}{-}
\DeclareUnicodeCharacter{2243}{$\simeq$}
\DeclareUnicodeCharacter{202F}{ }

\begin{document}
\title{Scale-dependence in $\Lambda$CDM parameters inferred from the CMB: a possible sign of Early Dark Energy}

\author{Jun-Qian Jiang}
\email{jiangjunqian21@mails.ucas.ac.cn}
\email{jiangjq2000@gmail.com}
\affiliation{School of Physical Sciences, University of Chinese Academy of Sciences, Beijing 100049, China}

\begin{abstract}
The early dark energy (EDE) model is one of the promising solutions to the Hubble tension.
One of the successes of the EDE model is that it can provide a similar fit to the $\Lambda$CDM model for the CMB power spectrum.
In this work, I analyze the phenomenology of the EDE and $\Lambda$CDM parameters on the CMB temperature power spectrum and notice that this cannot hold on all scales.
Thus, if the real cosmology is as described by the EDE model, the $\Lambda$CDM parameters will be scale-dependent when fitting the CMB power spectrum with the $\Lambda$CDM model, which can be hints for the EDE model.
I examine CMB-S4-like observations through mock data analysis and find that parameter shifts are notable.
As observations include smaller scales, I find lower $H_0$, $n_s$, $\omega_b$ and higher $\omega_m$, $A_s e^{-2\tau}$, which will also constitute new tensions with other observations.
They can serve as a possible signal for the EDE model.
\end{abstract}

\maketitle
\newpage

\section{Introduction}

The Hubble tension~\cite{Verde:2019ivm,DiValentino:2021izs,Perivolaropoulos:2021jda,Schoneberg:2021qvd,Shah:2021onj,Abdalla:2022yfr,DiValentino:2022fjm,Hu:2023jqc,Verde:2023lmm} is one of the greatest conflicts encountered in cosmological observations today.
The Hubble constant $H_0$ inferred from the CMB power spectrum observation prefers a low value.
For example, Planck 2018 reported $H_0 = (67.4\pm 0.5)$ km/s/Mpc~\cite{Planck:2018vyg}.
In contrast, local measurements, which are independent of the cosmological model, prefer a large value.
Based on observations of Cepheid-calibrated SNeIa, the SH0ES team reported $H_0 = (73.04 \pm 1.04)$ km/s/Mpc~\cite{Riess:2021jrx}.
Depending on different observations, there is a tension of about $4 \sim 6 \sigma$~\cite{Verde:2019ivm}.

This tension calls for some modification of the $\Lambda$CDM model.
These modifications to cosmology can usually be divided into post-recombination solutions and pre-combination solutions.
The former's capacity to resolve Hubble tension is limited by measurements from BAO and uncalibrated SNeIa~\cite{Bernal:2016gxb,Addison:2017fdm,Lemos:2018smw,Aylor:2018drw,Schoneberg:2019wmt,Knox:2019rjx,Arendse:2019hev,Efstathiou:2021ocp,Krishnan:2021dyb,Cai:2021weh,Keeley:2022ojz,Gomez-Valent:2023uof,Jiang:2024xnu},
which makes pre-recombination solutions more likely.
One of the promising solutions is a new pre-combination component named Early Dark Energy (see e.g. Refs.\cite{Karwal:2016vyq,Poulin:2018cxd,Kaloper:2019lpl,Agrawal:2019lmo,Lin:2019qug,Smith:2019ihp,Niedermann:2019olb,Sakstein:2019fmf,Ye:2020btb,Gogoi:2020qif,Braglia:2020bym,Lin:2020jcb,Odintsov:2020qzd,Seto:2021xua,Nojiri:2021dze,Karwal:2021vpk,Rezazadeh:2022lsf,Poulin:2023lkg,Odintsov:2023cli}),
although it also faced some challenges (see a recent review Ref.~\cite{Vagnozzi:2023nrq}).
{
In principle, for a given early dark energy model, it can be fitted directly to the observational data and can be detected by future CMB observations if the required new physics is indeed as it is described.
However, the early dark energy model is a class of models with numerous realizations that have not exactly the same phenomenologies.
These phenomenological differences will change their fit to CMB observations, which will ultimately lead to the fact that it may be difficult to detect by future CMB observations if we are not lucky enough to test the correct EDE model.
Actually, we have already seen in recent observations that current CMB data has different levels of preference for different EDE models~\cite{Poulin:2021bjr}.
Therefore, instead of fitting with a specific EDE model, a feasible approach is to observe the behavior when fitting with the \( \Lambda \)CDM model. At the same time, when detection capabilities are stronger, fitting with all features will inevitably introduce differences between different EDE models. Thus, to make the results more generalizable, it is necessary to compress these features into a few that are less sensitive to differences between EDE models, typically focusing on the relative relationships of the quantities rather than their specific values.
Although these compressed features cannot strictly prove the correctness of the EDE model, they can serve as a hint or a way to rule out the EDE model.
For example, the compressed features considered in Refs.~\cite{Farren:2021grl,Brieden:2022heh} are whether there is a significant difference between the sound horizon-dependent \( H_0 \) and the sound horizon-independent \( H_0 \).
Here, I consider the scale dependence of the $\Lambda$CDM parameters inferred from the CMB.
}

The scale inconsistency of the $\Lambda$CDM model, if it indeed exists, could advance our understanding of cosmological tensions and suggest the possibility of new physics~\cite{Akarsu:2024qiq}.
Actually, there appears to be some inconsistency in the fit of the $\Lambda$CDM to the current CMB observations, both within the same observations and across different observations.
For example, there is a $\gtrsim 2 \sigma$ discrepant between some parameters inferred from low-$\ell$ ($\ell < 800 (1000)$) and full-$\ell$ Planck 2015 TT power spectrum~\cite{Addison:2015wyg,Planck:2016tof}.
Meanwhile, the CMB lensing effect acting on the Planck CMB power spectrum also appears to be stronger than predicted by the $\Lambda$CDM model~\cite{Addison:2015wyg,Motloch:2019gux,Planck:2018vyg,Motloch:2018pjy} (see also Refs.\cite{DiValentino:2019dzu,DiValentino:2020hov,ACT:2020gnv,SPT-3G:2021wgf,SPT-3G:2022hvq}).
In addition, the Planck results also seem to have some discrepancies with the results of other CMB observations such as ACT~\cite{ACT:2020gnv} and SPT-3G~\cite{SPT-3G:2021wgf,SPT-3G:2022hvq}, see Refs.\cite{Handley:2020hdp,DiValentino:2022rdg,Forconi:2021que,Giare:2022rvg,DiValentino:2022oon} for discussions.

In this work, I investigate the role of the general EDE model and the $\Lambda$CDM parameters for the CMB power spectrum.
For simplicity, I restrict the analysis to the temperature (TT) power spectrum.
\footnote{
Another reason is that the polarization spectrum has a more significant dependence on the details of the EDE model implementation than the temperature spectrum, which is what I want to avoid here.
For example, it seems that the Planck polarization spectrum is related to the preference of a flat potential at high redshift for the axion-like EDE~\cite{Smith:2019ihp}, in contrast to the rock`n'roll model~\cite{Agrawal:2019lmo}.
Ref.\cite{Lin:2019qug} found that ADE and axion-like EDE have very similar fits for the TT spectrum, but exhibit different fits for the polarization spectrum.
}
I find that the degeneracy between the EDE model and the $\Lambda$CDM model becomes more and more imperfect as the range of scales covered by the CMB observations increases.
If our universe is as described by the EDE model, then as future CMB observations are able to cover smaller scales, a predictable result is that when the $\Lambda$CDM model is used to fit the observations, its parameters will shift, i.e., have scale dependencies.

This paper is organized as follows: In \autoref{sec:phe}, I discuss the main phenomenology of EDE on the temperature power spectrum and the reasons for the corresponding shifts in cosmological parameters during data fitting.
The results of the parameter shifts when fitting with the $\Lambda$CDM model are presented and discussed in \autoref{sec:fit}.
In \autoref{sec:ADE}, I demonstrate the robustness of the conclusions with respect to the details of the EDE model.
In addition, \autoref{sec:addpol} presents the results of parameter shifts when the polarization power spectrum is included in the analysis.

\section{EDE phenomenology and the shift of cosmological parameters} \label{sec:phe}

Early dark energy (EDE) is a component assumed to exist prior to the recombination that behaves like dark energy, i.e., with the equation of state $w_i \sim -1$.
Its energy fraction grows gradually and then decays before the recombination.
The observations favor the peak of the EDE energy fraction around the matter radiation equality, allowing for a maximum reduction of $r_s$~\cite{Poulin:2018cxd}.
The CMB observations also prefer that the energy decays faster than the radiation, i.e. the equation of state during the decay period $w_f \gtrsim 1/3$~\cite{Poulin:2018cxd,Lin:2019qug}.
There are many implementations that achieve the rapid decay of energy, such as oscillations~\cite{Poulin:2018cxd,Agrawal:2019lmo}, a runaway potential~\cite{Lin:2019qug}, anti-de Sitter (AdS) phase~\cite{Ye:2020btb,Ye:2020oix}, and phase transition~\cite{Niedermann:2019olb}.

Below I briefly summarize the general properties of EDE, its implications for the CMB TT power spectrum, and how the other $\Lambda$CDM parameters shift.
This means that I will not delve into the perturbative properties of EDE, although this is important for data fitting \cite{Karwal:2016vyq}. And there seems to be no qualitative difference in their implications for the CMB TT power spectrum between viable EDE models \cite{Poulin:2023lkg}. 

To aid understanding, I choose $\phi^4$ AdS-EDE~\cite{Ye:2020btb,Jiang:2021bab,Jiang:2022uyg} as an example.
I chose the best-fit points of AdS-EDE and $\Lambda$CDM for state-of-the-art CMB + BAO + SNeIa + $H_0$ observations.
Where the CMB dataset combination includes the Planck 2018 (PR3) \texttt{Commander} likelihood~\cite{Planck:2019nip} for the low-$\ell$ TT spectrum, the PR4 \texttt{lollipop} likelihood~\cite{Hamimeche:2008ai,Mangilli:2015xya,Tristram:2020wbi} for the low-$\ell$ EE spectrum, the PR4 \texttt{CamSpec} likelihood~\cite{Rosenberg:2022sdy} for the high-$\ell$ TTTEEE spectrum,
ACT DR4~\cite{ACT:2020gnv} and SPT-3G Y1~\cite{SPT-3G:2021wgf,SPT-3G:2022hvq} TTTEEE spectrum.
$\ell > 1800$ part of the ACT TT spectrum is removed to avoid the cross-correlation, following Ref.\cite{ACT:2020gnv}.
The BAO dataset consists of 6dF~\cite{Beutler:2011hx}, SDSS~\cite{Ross:2014qpa,BOSS:2016wmc,eBOSS:2020yzd}, and DESI Y1~\cite{DESI:2024mwx} datasets, following Ref.\cite{DESI:2024mwx}'s way of combining the latter two.
I also used the Pantheon+ measurement~\cite{Brout:2022vxf} on the SNeIa and the SH0ES data~\cite{Riess:2021jrx}.
The values of the parameters are shown in \autoref{tab:fid_parameters}.

\begin{table}[]
    \centering
    \begin{tabular}{|c|c c|} \hline \hline
        Parameters     & $\Lambda$CDM & EDE \\ \hline
        $\ln(1+z_c)$   &              & 8.07 \\
        $f_\text{EDE}$ &              & 0.11 \\
        $H_0$          & 68.33        & 72.22 \\
        $n_s$          & 0.9713       & 0.9913 \\
        $\omega_b=\Omega_b h^2$& 0.0223       & 0.0229 \\
        $\omega_c=\Omega_c h^2$& 0.1173       & 0.1343 \\
        $\ln(10^{10}A_s)$& 3.0529     & 3.0847 \\
        $\tau$         & 0.0592       & 0.0552 \\ \hline
        $\omega_m=\Omega_m h^2$     & 0.1403       & 0.1578 \\
        $10^9 A_\mathrm{s} e^{-2\tau}$ & 1.8811 & 1.9576 \\ \hline
        $100\,\theta_s^*$& 1.0446       & 1.0437 \\
        $100\,\theta_D$  & 0.3232       & 0.3262 \\
        $\text{PH}_1 / \text{PH}_2$     & 2.197 & 2.193 \\
        $C^{\phi\phi}_{\text{peak}}$    & $1.360\cross 10^7$ & $1.395\cross 10^7$ \\
        $C^\text{TT, EISW}_\text{peak}$ & 409.0 & 405.5 \\ \hline \hline
    \end{tabular}
    \caption{The best-fit values of the current observations (described in the text) for parameters of the two models.}
    \label{tab:fid_parameters}
\end{table}

\begin{table}[h]
    \centering
    \begin{tabular}{|c|c c c c c c|} \hline \hline
                                        &  EDE      &  $H_0$                &  $\omega_m$            &  $n_s$                &  $\omega_b$            & $A_s \exp(-2\tau)$\\ \hline
        $\theta_s^*$                    &  $-2.6\%$ &  $+1.1\% = 0.2\Delta$ & $+1.6\% = 0.13\Delta$  &  $0$                  & $-0.2\% = -0.08\Delta$ & $0$ \\
        $\theta_D$                      &  $-0.9\%$ &  $+1.1\% = 0.2\Delta$ & $+1.4\% = 0.12\Delta$  &  $0$                  & $-0.6\% = -0.3\Delta$  & $0$ \\
        $\text{PH}_1 / \text{PH}_2$     &  $+0.7\%$ &  $0$                  & $-0.6\% = -0.05\Delta$ & $-1.8\% = 0.9\Delta$  & $+1.5\% = 0.7\Delta$   & $0$\\
        $C^{\phi\phi}_{\text{peak}}$    &  $-3.1\%$ &  $-2.2\% = -0.4\Delta$& $+8.2\% = 0.7\Delta$   & $-3.1\% = 1.5\Delta$  & $-0.2\% = -0.08\Delta$ & $+3.2\% = 1\Delta$ \\
        $C^\text{TT, EISW}_\text{peak}$ & $+16.0\%$ &  $0$                  & $-15.8\% = -1.3\Delta$ & $-3.0\% = -1.5\Delta$ & $+0.1\% = 0.05\Delta$  & $+3.2\% = 1\Delta$ \\ \hline \hline
    \end{tabular}
    \caption{The fractional change to the quantities in the first column produced by respectively adding the EDE or varying the $\Lambda$CDM parameter from the $\Lambda$CDM model to the EDE model. $\Delta$ stands for the fractional change of the parameters $x$: $\Delta = (x_\text{EDE} - x_{\Lambda\text{CDM}})/ x$.}
    \label{tab:responds}
\end{table}

Here, I use five quantities to characterize the effect of different components/parameters on the CMB TT power spectrum:
\begin{itemize}
    \item The angular separation of acoustic peaks $\theta_s^* = r_s^* / D_A$, where $D_A$ is the angular distance to the last scattering surface and $r_s^*$ is the sound horizon at CMB last-scattering: \begin{equation}
        r_s^* = \int_{z^*}^\infty \frac{c_s(z)}{H(z)} \dd z \, ,
    \end{equation}
    $z_*$ is the redshift to the CMB last-scattering surface\footnote{A better way is to use visibility-averaged sound horizon~\cite{Knox:2019rjx}, same as for the $r_D$ below.} and $c_s$ is the sound speed.
    \item The angular size $\theta_D = r_D / D_A$ of the diffusion damping scale $r_D$.
    \item The ratio between the first and second acoustic peak $\text{PH}_1 / \text{PH}_2$. It is related to the shift of the zero point of the acoustic oscillation.
    {
    Meanwhile, since the position of the first peak is close to the project sound horizon at the matter-radiation equality, it will also be related to the “potential envelope” \cite{Hu:1996vq,Hu:1996mn,Hu:1995kot}.
    This effect arises from the decay of the potential energy during and before the matter-radiation equality, which leads to a near-resonant driving of the modes entering the horizon.
    It leads to an exponential dependence of the CMB power spectrum if the diffusion damping is removed~\cite{Hu:1996mn}:
    \begin{equation}
        1 + A \exp(-1.4 \ell_\text{eq}/\ell),
    \end{equation}
    where the matter-radiation equality $\ell_s^\text{eq} \simeq 223$~\cite{Knox:2019rjx}.
    }
    I do not use $\ell_s^\text{eq}$ directly below because it cannot fully capture the effects of EDE.
    \item CMB lensing will smooth the peak and increase the small-scale power, which is the dominant contribution at very small scales. I use the peak of the power spectrum of the lensing potential $C^{\phi\phi}_{\text{peak}}$ to characterize the strength of this effect.
    \item The early integrated Sachs–Wolfe (EISW) effect~\cite{Sachs:1967er} is important for the scale around and larger than the first acoustic peak.
    {
    Even at the last scattering surface, it is not completely matter-dominated, and the radiation is still causing the potential to decay.
    As a result, the energy of the photon at the corresponding scale is altered.
    }
    I use the peak of the TT power spectrum contributed solely from the EISW effect $C^\text{TT, EISW}_\text{peak}$ to characterize it. Actually, it has been found both $\Lambda$CDM and the EDE model can lead to a consistent EISW effect \cite{Vagnozzi:2021gjh}.
\end{itemize}
In \autoref{tab:responds}, I show the effect of different components/parameters on these quantities by changing these components/parameters from the $\Lambda$CDM best fit to the AdS-EDE best fit, respectively.

\begin{figure}
    \centering
    \includegraphics[width=\textwidth]{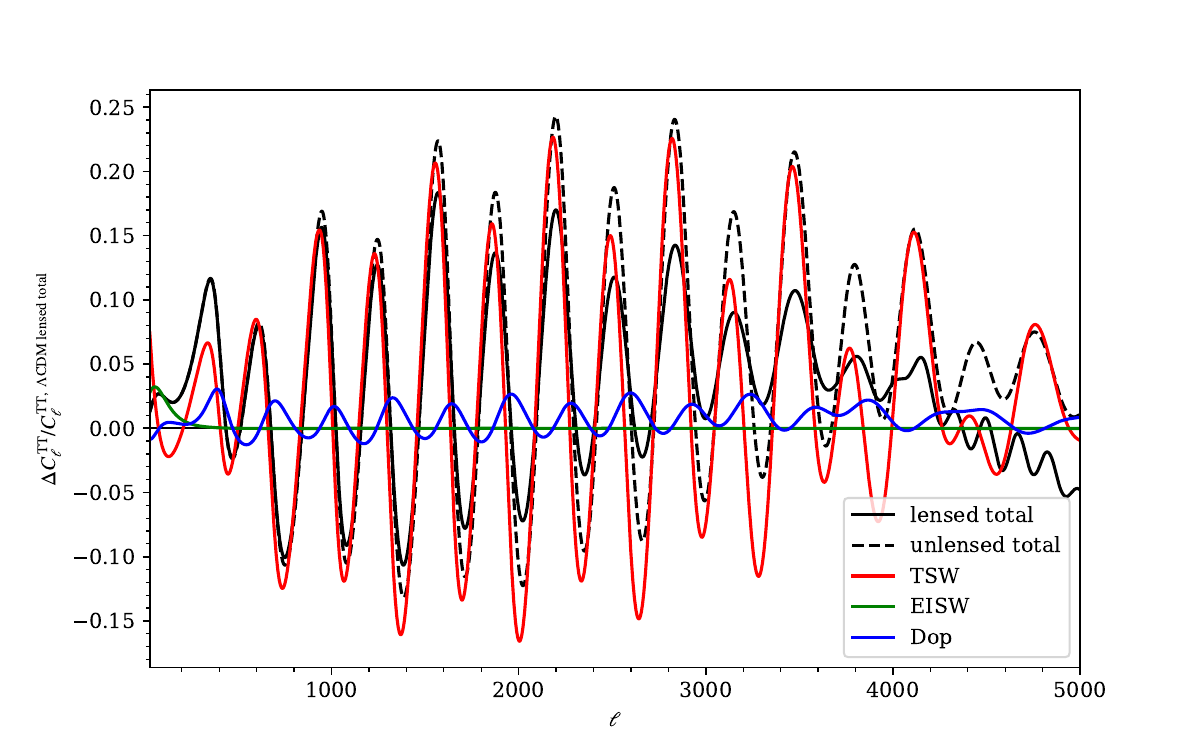}
    \caption{The effect on the TT power spectrum of adding the EDE (the best-fit value of the EDE model) to the best-fit $\Lambda$CDM model.
    {
    The colored lines are the effect on the unlensed TT spectrum where only a certain contribution is considered. “TSW” means intrinsic temperature corrected by Sachs-Wolfe, i.e. the contribution that comes from the difference between the potential at horizon crossing and today.
    “EISW” means early ISW effect, as explained above.
    And “Dop” means the Doppler effect, i.e. the contribution from Doppler shifts due to out-of-phase oscillations of the fluid with respect to temperature., which can be seen from the figure that it is always a minor contribution compared to TSW and EISW.
    }}
    \label{fig:temperature_contributions}
\end{figure}

Regardless of the implementation, all the EDE models provide a higher background expansion rate $H(z)$ in a period before and around the recombination, which leads to a smaller sound horizon $r_s^*$.
Since the EDE itself does not significantly alter the late expansion history, the angular distance $D_A$ to the last scattering surface is unchanged, which leads to a reduction in the angular separation of acoustic peaks $\theta_s^* = r_s^* / D_A$.
Therefore, the project sound horizon $\ell_s \sim \pi / \theta_s^*$ is increased, which means the shift of the sound peak in the power spectrum.
This leads to the most significant oscillating residuals in \autoref{fig:temperature_contributions}.

\begin{figure}
    \centering
    \includegraphics[width=\textwidth]{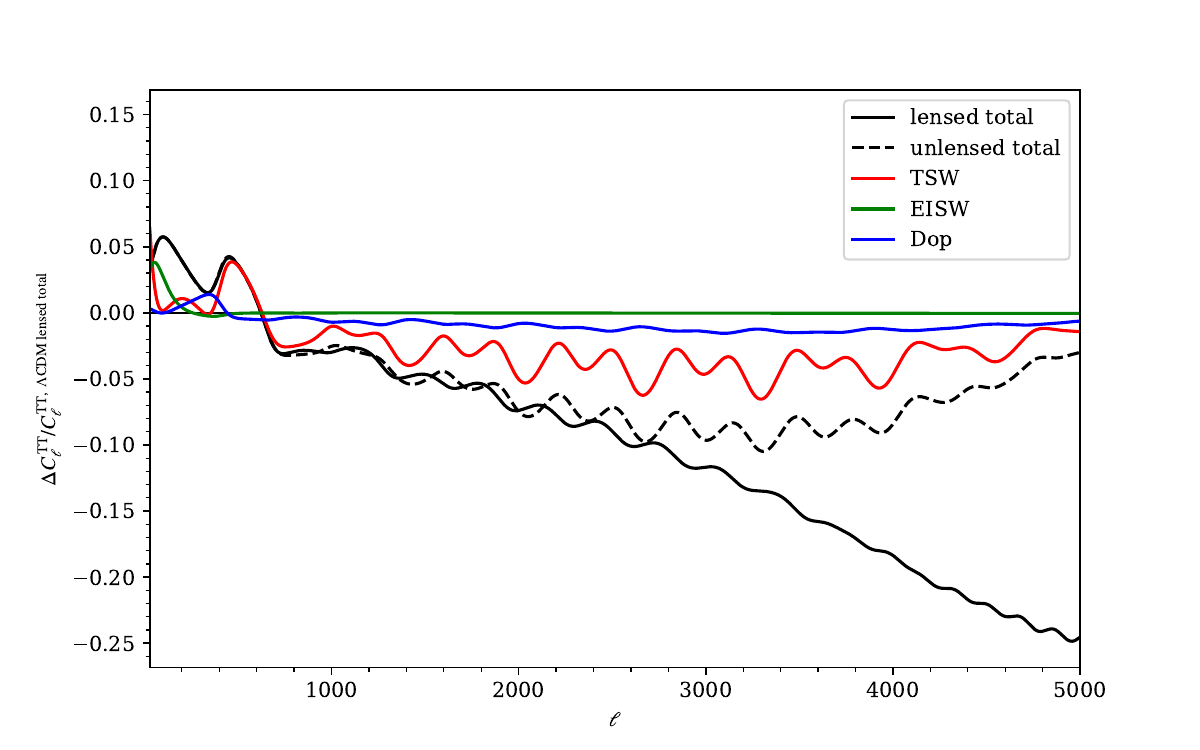}
    \caption{Similar to \autoref{fig:temperature_contributions}, but fixed $\theta_s^*$ by raising $H_0$.}
    \label{fig:temperature_contributions_fixed_theta}
\end{figure}

Meanwhile, the higher expansion rate $H(z)$ around the recombination also reduces the diffusion damping scale $r_D$.
Similarly, $\theta_D = r_D / D_A$ is reduced and $\ell_D \sim \pi / \theta_D$ is raised.
Therefore, the diffusion damping is suppressed and the amplitude of the CMB spectrum at high $\ell$ is raised.

{
Another implication of the higher expansion rate $H(z)$ is a faster decay of the Weyl potential since EDE accelerates the expansion of the early universe at a period before recombination and we will get a shallower Weyl potential eventually, see e.g. figures in \cite{Jiang:2022uyg,Lin:2019qug,Niedermann:2020dwg,Poulin:2023lkg}.
It affects the CMB power spectrum through three effects.
The first two are the enhanced “potential envelope” and the early ISW effect as they arise from the decay of the potential, which is enhanced in the case of EDE.
}
As a result, the $\text{PH}_1 / \text{PH}_2$ and $C^\text{TT, EISW}_\text{peak}$ gets higher. 
On the other hand, a shallower Weyl potential also implies a lower matter power spectrum in the late universe.
This also means that the lensing effect is weaker in the CMB TT power spectrum.
In particular, it significantly reduces the amplitude at very high $\ell$ where CMB lensing is the dominant effect.

\begin{figure}
    \centering
    \includegraphics[width=\textwidth]{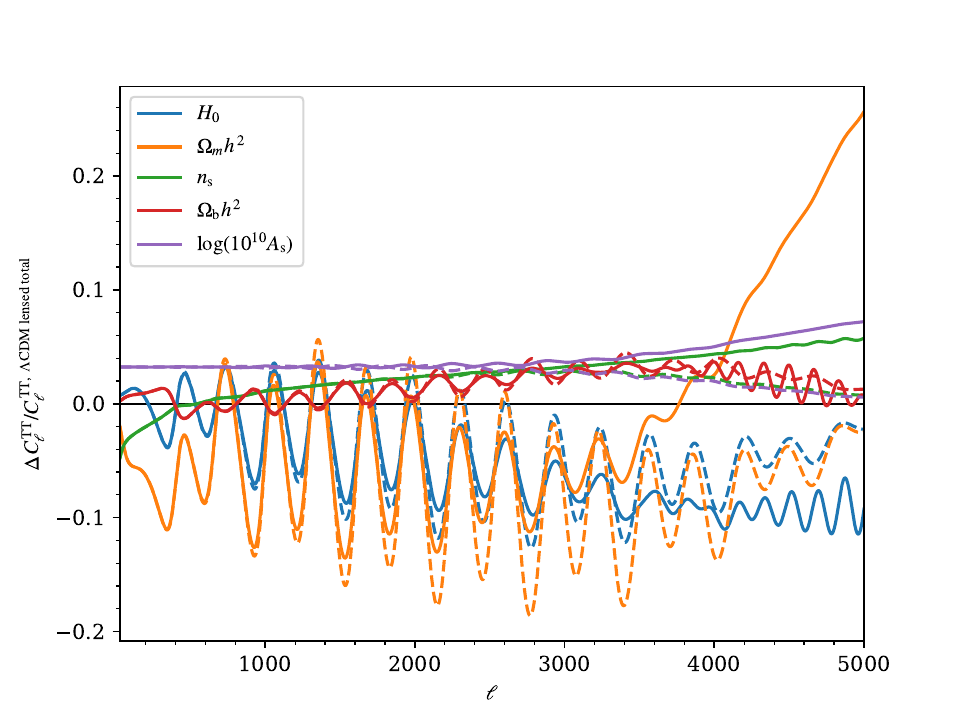}
    \caption{The effect on the TT power spectrum of varying the $\Lambda$CDM parameter from the $\Lambda$CDM model to the EDE model respectively. The solid lines show the effect on the lensed spectrum while the dashed lines show the effect on the unlensed spectrum.}
    \label{fig:responds_LCDM}
\end{figure}

When we fit the EDE model to CMB observations, the other parameters will be shifted to reconcile the CMB power spectrum.
Since we know that the $\Lambda$CDM model explains the observed CMB power spectrum well, we usually expect that by shifting the parameters, the EDE model produces a power spectrum similar to that of the $\Lambda$CDM model.
However, this may not always be true since we have actually found that for some combinations of CMB datasets, the EDE model can fit the observations better than the $\Lambda$CDM model (e.g.~\cite{Hill:2021yec,Poulin:2021bjr,LaPosta:2021pgm,Smith:2022hwi,Jiang:2022uyg,Peng:2023bik,Murgia:2020ryi,Jiang:2023bsz}).
\footnote{However, for some CMB datasets, the EDE model may not fit the CMB observations better than the $\Lambda$CDM model if the local $H_0$ measurements are not fitted simultaneously, especially given the number of parameters added by EDE (e.g.~\cite{Efstathiou:2023fbn}).}
Nevertheless, we still assume in this section that the EDE model tries to fit a $\Lambda$CDM-like power spectrum because it is not yet clear whether this preference from CMB for the EDE model is general.

The most obviously shifted parameter is $H_0$, as it will reconcile the residuals due to the significant acoustic peak shift by changing the angular distance
\begin{equation} \label{eq:DAint}
    D_A = 2998\,\text{Mpc} \times \int \frac{\dd z}{\sqrt{\omega_m(1+z)^3 + (H_0^2 - \omega_m)}} \, .
\end{equation} 
This is the direct reason why the Hubble tension has been relieved.
It simultaneously increases $\theta_s^*$ and $\theta_D$.
A large $H_0$ also reduced the lensing effect, partly due to the reduction in the distance the light travels after deflection and the change in the $k$-$L$ projection relation.

Another parameter that can significantly affect the positions of the acoustic peaks is $\omega_m$, which can change $\theta_s^*$ by affecting the acoustic horizon $\delta r_s^* / r_s^* \sim 1/4\, \delta \omega_m / \omega_m$.
However, as shown in \autoref{eq:DAint}, it will also reduce $D_A$ and finally result in a larger $\theta_s^*$.
Its effect on $\theta_D$ is similar but weaker, which can be clearly seen in \autoref{fig:temperature_contributions_fixed_theta}.
Meanwhile, the reduced decay of Weyl potential, as opposed to the EDE (but with a different level of response), also helps reduce the “potential envelope” and the early ISW effect.
These effects
result in a lower $\text{PH}_1 / \text{PH}_2$.
Finally, a larger growth of matter perturbations leads to a stronger lensing effect.
In some work, it was argued that BAO and uncalibrated SNeIa can constrain $\Omega_m$ because the late universe is still $\Lambda$CDM, and thus $\omega_m$ is lifted with $H_0$~\cite{Knox:2019rjx,Ye:2020oix,Pogosian:2020ded,Jedamzik:2020zmd,Poulin:2024ken,Pedrotti:2024kpn}.
In this work, I do not rely on this point as I only consider the CMB power spectrum.
However, it is interesting that the present constraints on the EDE model obtained a $\Omega_m$ consistent with the results of BAO and uncalibrated SNeIa.
We will see in the next section that this may be a coincidence due to the weak observation constraints currently (see also Figure 6 of Ref.\cite{Jiang:2021bab}).

A large $\omega_b$ can reduce both the sound horizon and the diffusion damping scale (but it's more effective for the latter), leading to lower $\theta_s^*$ and $\theta_D$.
The main effect of $\omega_b$ is altering the zero-point of the acoustic oscillations, resulting in a higher $\text{PH}_1 / \text{PH}_2$.

There are three reasons cause the raise of $n_s$.
Firstly, although $n_s$ does not affect $\theta_D$, It provides greater amplitude at small scales, thus counteracting the enhanced diffusion damping~\cite{Knox:2019rjx,Niedermann:2020dwg,Ye:2021nej,Jiang:2022uyg}.
This is required because the effect of the EDE on $\theta_s^*$ and $\theta_D$ is unbalanced, whereas shifts in the other parameters will either only change them equally or have very little effect.
Secondly, $n_s$ can directly balance $\text{PH}_1 / \text{PH}_2$~\cite{Ye:2021nej}.
Thirdly, $n_s$ will reduce the peak of the matter power spectrum as its scale is smaller than the pivot scale.
However, $n_s$ enhances $C^{\phi\phi}_L$ at high $L$, which can help to transfer higher power from large ($\ell - L$) to small ($\ell$) scales.
Therefore, the amplitude at the very high $\ell$ becomes higher instead to balance the effect of EDE.
Besides, $C^\text{TT, EISW}_\text{peak}$ gets lower as its scale is also smaller than the pivot scale.
Actually, an interesting fact is that if $H_0$ were to reconcile with the result of SH0ES, we would probably need a Harrison-Zeldovich spectrum $n_s=1$~\cite{Ye:2021nej,Jiang:2022uyg,Cruz:2022oqk,Jiang:2022qlj,Jiang:2023bsz,Peng:2023bik,Wang:2024tjd,Wang:2024dka}, which has a great impact on the inflation model (e.g. Refs.~\cite{Kallosh:2022ggf,Ye:2022efx}).

Finally, the role of $A_s \exp(-2\tau)$ is simple.
It directly changes the overall amplitude of the unlensed TT power spectrum and the lensing potential spectrum (since I vary $A_s$ here).
The latter leads to the further enhancement of the lensed TT power spectrum at very high $\ell$.
On the other hand, if one chooses to vary $\tau$, there is only the former effect.

\section{Scale-dependent $\Lambda$CDM parameters inferred from the CMB} \label{sec:fit}

While the effects of the component/parameter shifts mentioned above can in practice allow the EDE model to fit the CMB observations similar to $\Lambda$CDM, this is most likely a coincidence due to the current lack of observational precision as well as the limited range of scales.
Here are three main reasons why the degeneration of these two models cannot be satisfied at all scales:
\begin{itemize}
    \item The effect of EDE on the Weyl potential is not trivial, and it cannot be balanced simply by varying $\omega_m$.
    Therefore, scale-dependent “potential envelope” and EISW effect differences must exist between EDE and $\Lambda$CDM.
    This was brought up in Ref.\cite{Knox:2019rjx}, where they also suspected it may be related to the high-low $\ell$ inconsistencies~\cite{Addison:2015wyg,Planck:2016tof} and the lensing anomaly~\cite{Addison:2015wyg,Motloch:2019gux,Planck:2018vyg,Motloch:2018pjy} in the Planck 2018 data.
    However, the lensing anomaly seems to be not related to EDE \cite{Jiang:2022uyg,Fondi:2022xsf} and the lensing anomaly is reduced in the updated PR4 data~\cite{Tristram:2023haj,Rosenberg:2022sdy}.
    On the other hand, the evolution of the Weyl potential is closely related to the details of the EDE model.
    Hence I will not rely on this point here.
    \item In the above, I used 5 quantities to characterize the CMB TT power spectrum.
    Even considering the compressed information, $\Lambda$CDM model has fewer free parameters \footnote{$A_s \exp(-2\tau)$ is used to absorb the overall amplitude which is not mentioned above, while, as stated below, I consider the information from the polarization spectrum for $\tau$.}.
    Therefore it is not possible that all quantities can be well reconciled.
    Since these quantities are important at different scales, its back effect is the variation of $\Lambda$CDM parameters at different scales.
    \item Another important source of scale dependence is similar effects in the power spectrum but with different shapes as functions of scale, which I will elaborate on below.
\end{itemize}

\begin{figure}
    \centering
    \includegraphics[width=0.47\linewidth]{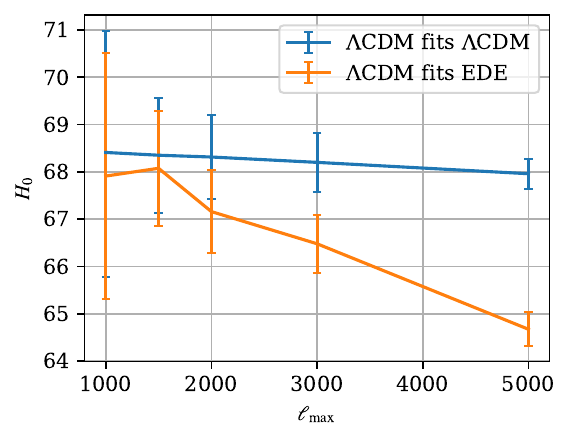}
    \includegraphics[width=0.47\linewidth]{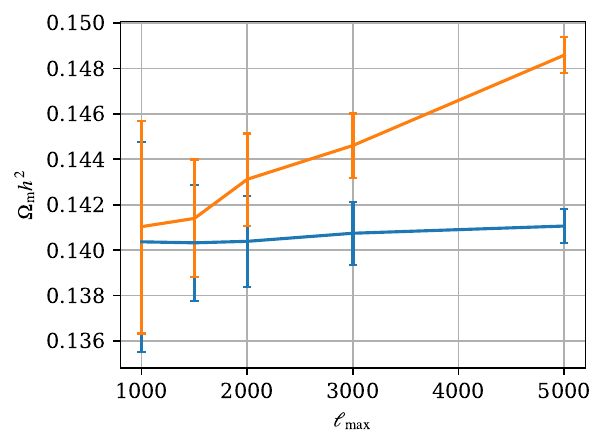}
    \includegraphics[width=0.47\linewidth]{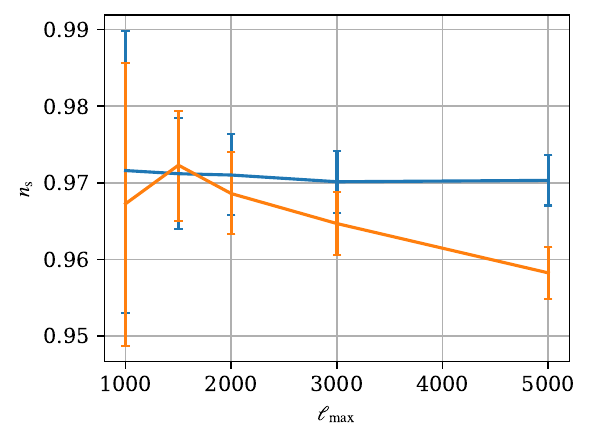}
    \includegraphics[width=0.47\linewidth]{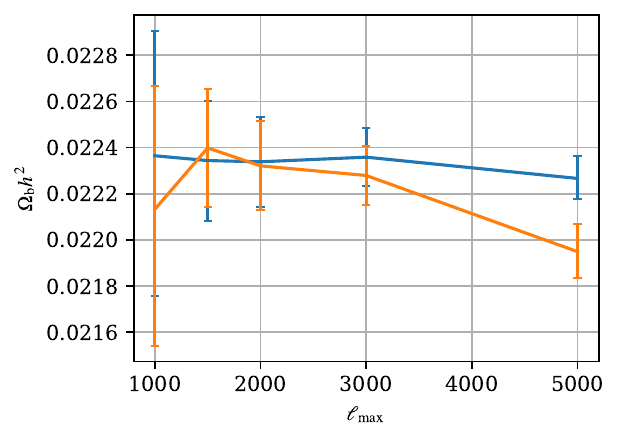}
    \includegraphics[width=0.47\linewidth]{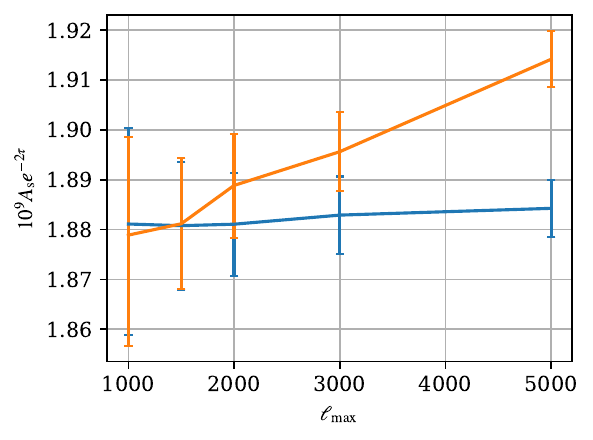}
    \includegraphics[width=0.47\linewidth]{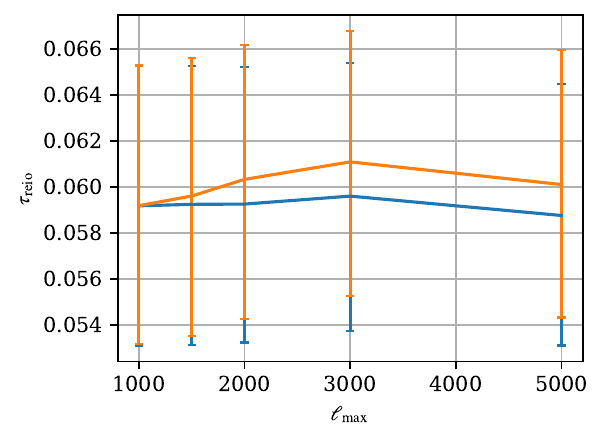}
    \caption{The posterior distribution of $\Lambda$CDM parameters inferred from mock CMB data with different cut-off $\ell_\text{max}$. The error bars show the mean and 68\% confidence intervals.}
    \label{fig:trends}
\end{figure}

I assume that the real universe is $\Lambda$CDM or AdS-EDE with parameter values as in \autoref{tab:fid_parameters}, and generate mock data using \texttt{cobaya\_mock\_cmb}\footnote{\url{https://github.com/misharash/cobaya_mock_cmb}. I used the \texttt{deproj0} noise curves.}\cite{Rashkovetskyi:2021rwg} for a CMB-S4-like observation.
Then I fit these two mock data till some $\ell_\text{max}$ respectively with the $\Lambda$CDM model.
Only the TT power spectrum is used, so I impose a prior for $\tau$ from the previously mentioned $\Lambda$CDM fit the current observation \footnote{While this may in principle introduce a bias for the case of EDE as mock data, it can be seen from the posterior of $\tau$ in \autoref{fig:trends} that $\tau$ does not have a significant effect.}.
I used \texttt{nautilus}~\cite{Lange:2023ydq} for importance nested sampling accelerated with machine learning.
The trends of the posterior distributions of the $\Lambda$CDM parameters varying with $\ell_\text{max}$ are shown in \autoref{fig:trends}.

We can find that the posterior distribution of $\ell_\text{max} \lesssim 1500$ is similar for both cases.
This is understandable, since the values of the mock data are taken from current observations on the same CMB, and the main constraining power of the current observations is at these scales.
When $\Lambda$CDM is used to fit a $\Lambda$CDM universe, there is no observable scale dependence of the parameters, which verifies that our pipeline has no bias.
In contrast, when $\Lambda$CDM is used to fit an EDE universe, the $\Lambda$CDM parameters, except $\tau$, produce a significant scale-dependent shift.
The largest shift lies between the full range ($\ell_\text{max} = 5000$) and $\ell_\text{max} \sim 1500$, where the parameters have a shift of about $1 \sim 2 \, \sigma$.

\begin{figure}
    \centering
    \includegraphics[width=\linewidth]{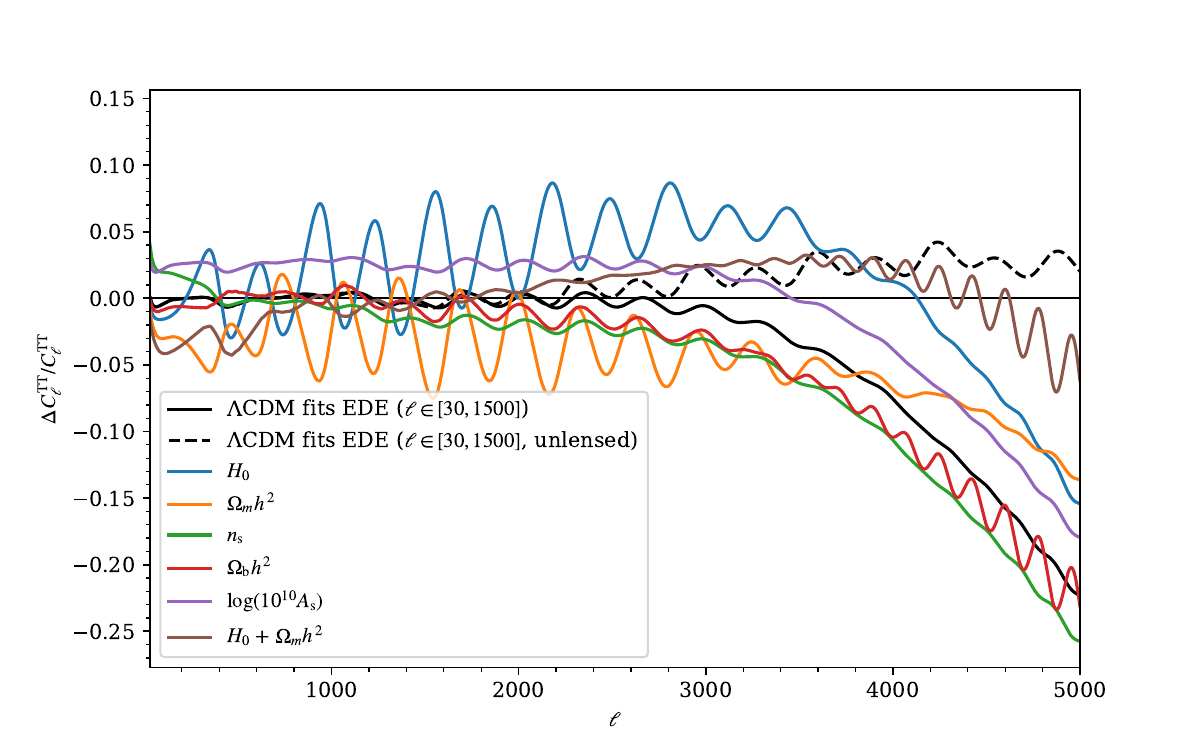}
    \caption{The black lines are the residual at the best-fit $\Lambda$CDM point ($\ell \in [30, 1500]$) of the TT power spectrum compared to mock data (EDE). The colored lines are the residuals after varying the parameters from the best-fit values of $\ell \in [30, 1500]$ to the best-fit values of $\ell \in [30, 5000]$, respectively. They are the residuals between lensed power spectrums when not stated.}
    \label{fig:compare_bestfit_LCDM_fit_EDE}
\end{figure}

To understand these behaviors, I show in \autoref{fig:compare_bestfit_LCDM_fit_EDE} the residuals of $\ell_\text{max} = 1500$ relative to the EDE mock data, and how the $\Lambda$CDM parameters shift to affect the residuals separately.
Firstly, at high $\ell$, $\Lambda$CDM has positive unlensed residuals, which is, in part, due to the mismatch between the primordial power spectrum as a power-law spectrum and the exponential suppression of the diffusion damping.
The fractional change to the unlensed TT power spectrum due to $n_s$ is proportional to $\ln(\ell/\ell_\text{pivot})$, while for $\theta_D$, it $\propto -\ell^2$.
Hence, even if they can cancel out at low $\ell$, they will inevitably diverge at high $\ell$.
This effect may be observable if we can delensing efficiently.
However, at very high $\ell$, the transfer from large to small scales due to CMB lensing is the dominant effect.
In \autoref{fig:responds_LCDM}, we can see that a large $\omega_m$ will induce a dramatic enhancement of the lensed TT spectrum at very high $\ell$.
When $\Lambda$CDM only fits the part of $\ell < 1500$, this effect is ignored and the constraints on $\omega_m$ will come mainly from the “potential envelope” and EISW effects, which favor a low $\omega_m$ due to the lack of EDE to produce a shallower Weyl potential.
As a result, this leads to large negative residuals in the unfitted part.
Actually, this is visible in the residuals of some EDE fits to the real observations (e.g. Figure 12 of Ref.\cite{Jiang:2021bab}, where instead the residuals of EDE relative to $\Lambda$CDM are shown.).

When $\ell_\text{max}$ moves to small scales, $\omega_m$ increases and $H_0$ decreases to compensate for the lensing effect.
At the same time, they maintain their relation to preserve the acoustic peak separation.
Their net effects, shown as brown lines, are negative residuals at large scales and positive residuals at $2000\lesssim \ell \lesssim 4000$.
The former, due to the reduced “potential envelope” and EISW effects, can be compensated by increasing the primordial power spectrum (lowering $n_s$ and increasing $A_s$).
The lower $n_s$ simultaneously mitigates the positive residuals at $2000\lesssim \ell \lesssim 4000$, while the enhancement of diffusion damping by lowering $\omega_b$ has a similar effect.

Moreover, since the parameters derived from the analysis using the full scale have large departures from $\ell_\text{max} \sim 1500$, whose results are similar to the current observations, the parameters based on the full-scale analysis will deviate from the current observations.
For example, the full-scale analysis finds $\Omega_m = 0.355 \pm 0.006$, which conflicts with the current BAO and SNeIa measurements.
Meanwhile, it finds $\omega_b = 0.0219 \pm 0.0001$, lower than the current constraint determined from BBN.

\section{Conclusion}

I analyzed the phenomenology for general EDE and parameters of $\Lambda$CDM.
I've found that because the effects of different components/parameters vary at different scales, it's not possible to keep $\Lambda$CDM degenerate with EDE at all scales.
Consequently, if the real universe is as described by the EDE model, then when we use $\Lambda$CDM to fit it, we will find that its parameters are scale-dependent.
Here I use the CMB-S4 observation on the TT power spectrum (with a prior on $\tau$ from the current observations) as an example to test this behavior.
As the range of the analysis increases from $\ell_\text{max} \sim 1500$ to small scales, $H_0$, $n_s$, and $\omega_b$ decreases, while $\omega_m$ and $A_s e^{-2\tau}$ increases.
Actually, if there is some evidence for non-zero parameters used to quantify the running of these parameters, such as the running of scalar spectral index $\alpha_s$, it could also be a sign of EDE.
The main reason for this scale dependence is that variations in the Weyl potential by the EDE will cause different effects between low $\ell$ and very high $\ell$.

{
Therefore, if the constraints on the $\Lambda$ CDM parameters in future CMB observations, e.g., CMB-S4~\cite{CMB-S4:2016ple}, Simons Observatory~\cite{SimonsObservatory:2018koc}, show the same scale-dependent shift direction at $\ell \gtrsim 1500$ as predicted here, then this is a possible signal for the EDE model as a solution to the Hubble tension.
Meanwhile, this hint relies on the phenomenology shared by the EDE models without delving into the differences between the EDE models, making it an important addition to the direct fitting of specific EDE models.
}
Moreover, my results also imply that if the real universe is as described by the EDE, some observables inferred from future CMB observations (assuming $\Lambda$CDM) will be in conflict with other observations. For example, a very large $\Omega_m$ could conflict with BAO and SNeIa observation which are independent from the early Universe.
Meanwhile, the constraints for $\omega_b$ are lower than those from BBN.
These suggest that new tensions may arise.

\begin{acknowledgments}
I would like to thank Yun-Song Piao, Gen Ye, William Giarè, and Sunny Vagnozzi for their helpful discussions. 
I acknowledge the use of high performance computing services provided by the International Centre for Theoretical Physics Asia-Pacific cluster and Scientific Computing Center of University of Chinese Academy of Sciences.
\end{acknowledgments}

\appendix

\section{Scale-dependent $\Lambda$CDM parameters with various mock ADE models} \label{sec:ADE}

Here, I investigate the universality of the above conclusion for the EDE model through various Acoustic Dark Energy (ADE) models~\cite{Lin:2019qug}.
In the ADE model, EDE is simulated as a perfect fluid.
The equation of state of ADE is modeled as:
\begin{equation}
    1 + w(a) = \frac{1+w_f}{[1 + (a_c/a)^{3(1+w_f)/p}]^p} \, .
\end{equation}
Around the critical scale factor \( a_c \), \( w \) transitions from \( -1 \) to \( w_f \), and the rapidity of the transition is controlled by \( p \).
The fractional energy density of ADE at $a_c$ is denoted as $f_\text{EDE}$.
In addition, there is another adjustable parameter, which is the sound speed \( c_s^2 \) of the ADE.
With these five parameters, ADE can mimic the behavior of many other EDE models, especially in terms of the background dynamics of interest here.
For example, the AdS-EDE model tested above corresponds to \( c_s^2 = 1 \) and has \( w_f \gtrsim 1 \) in the AdS phase.
The axion-like EDE model (\( n = 3 \)) has \( w_f = 1/2 \), \( p = 1 \), and \( c_s^2 \in [1/2, 1] \).
The canonical ADE in Ref.\cite{Lin:2019qug} has \( w_f = 1 \) and \( c_s^2 = 1 \) fixed.

I selected a toy ADE model ($z_c=3400, c_s^2=0.75, p=1, w_f=0.75$) as the fiducial model, and then for each ADE parameter, I varied it within a reasonable range to evaluate its impact on the conclusion mentioned above, namely, the scale dependence of the $\Lambda$CDM parameters inferred from the CMB.
The range for each parameter is as follows:
\begin{align}
    z_c &= [3000, 4000]\\
    c_s^2 &= [0.5, 1]\\
    p &= [0.5, 2]\\
    w_f &= [0.5, 1]
\end{align}
The ADE fractional energy density \( f_\text{EDE} \) at \( z_c \) is always fixed at $0.1$, because the effect of EDE should be simply proportional to \( f_\text{EDE} \) at leading order.
It is important to note that the ADE's \( f_\text{EDE} \) is not equal to the peak value of \( f_\text{EDE}(z) \), so its specific value should not be directly compared with \( f_\text{EDE} \) in other EDE models.
At the same time, other ADE parameters also affect the relationship between \( f_\text{EDE} \) and the peak value of \( f_\text{EDE}(z) \) (especially the parameter \( p \)), so they also influence the strength of the ADE effect.
As a result, only the direction of the parameter shifts should be compared, rather than the magnitude of the shifts.
The other cosmological parameters are fixed as follows:
\begin{equation}
    H_0 = 73.04, \ n_s = 1, \ \Omega_bh^2 = 0.02218, \ \Omega_m = 0.3, \ \ln(10^{10}A_s) = 3.08, \ \tau = 0.055 \, .
\end{equation}

\begin{figure}
    \centering
    \includegraphics[width=\linewidth]{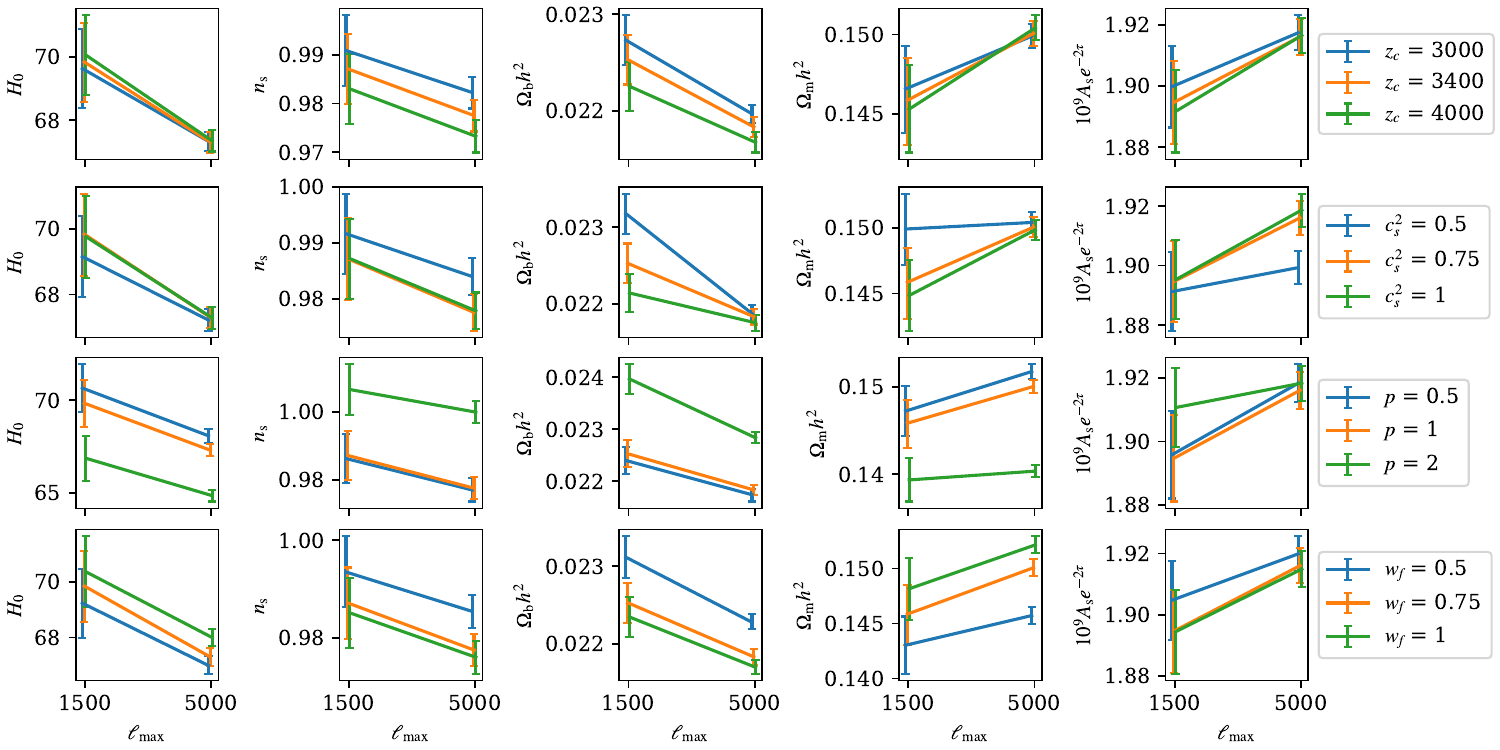}
    \caption{The posterior distribution of the $\Lambda$CDM parameters inferred from mock CMB data with different cut-off \( \ell_\text{max} \).
    For each row, the assumed mock data is generated by altering one parameter based on the fiducial ADE model (\( z_c = 3400 \), \( c_s^2 = 0.75 \), \( p = 1 \), \( w_f = 0.75 \)) as shown in the legend on the right.
    The error bars show the mean and 68\% confidence intervals.}
    \label{fig:trend_ADE}
\end{figure}

The results of the parameter shifts are shown in \autoref{fig:trend_ADE}, which demonstrate that the decrease in \( H_0 \), \( n_s \), and \( \omega_b h^2 \) with increasing \( \ell_\text{max} \) is sufficiently robust with respect to various ADE parameters.
For the parameters \( \omega_m \) and \( A_s e^{-2\tau} \), although the shifts in the parameters are not as pronounced in some edge cases, the direction of the shifts remains consistent with the original conclusion.

\section{Scale-dependent $\Lambda$CDM parameters inferred from the CMB including the polarization spectrum} \label{sec:addpol}

\begin{figure}
    \centering
    \includegraphics[width=\linewidth]{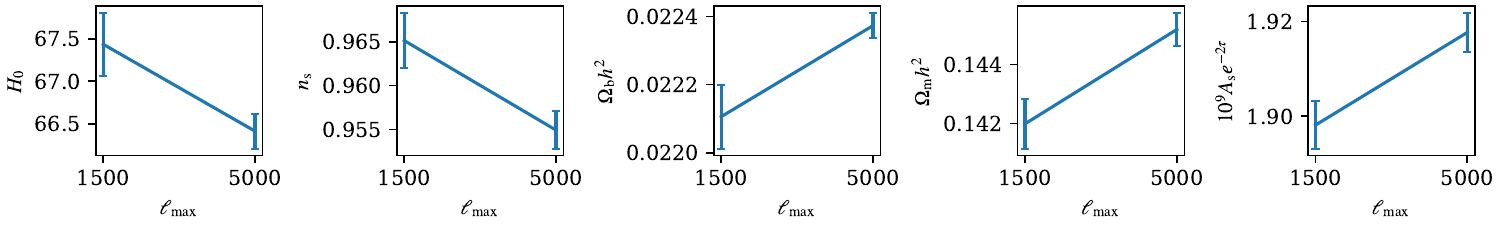}
    \caption{The posterior distribution of the $\Lambda$CDM parameters inferred from mock CMB data with different cut-off \( \ell_\text{max} \).
    The mock model is the same as the one used in \autoref{fig:trends} but the polarization spectrum (TE, EE) is also included in the mock CMB data.
    The error bars show the mean and 68\% confidence intervals.}
    \label{fig:add_pol}
\end{figure}

In \autoref{fig:add_pol}, I present the results with the inclusion of the polarization spectrum in the fitting, where the conclusion changes, and \( \omega_b h^2 \) increases with \( \ell_\text{max} \) instead.
This is understandable because the above analysis was conducted with respect to the temperature spectrum.
Since different EDE models exhibit significant differences in the polarization spectrum, the trends shown here may not be applicable to all EDE models.

\bibliography{main}
\end{document}